\documentclass[a4paper,10pt]{article}
\usepackage[utf8]{inputenc}
\usepackage{fullpage}
\usepackage{hyperref}
\usepackage{amsmath}
\usepackage{graphicx}
\usepackage{authblk}
\usepackage{amsfonts}

\newcommand{\RN}{{{\bf{R}}^N}}
\newcommand{\bg}{{{\bf{g}}}}
 
\newcommand{\intpmi}{{{\int_{-\infty}^{+\infty}}}}

\newtheorem{theorem}{Theorem}

\begin{document}

\title{Convergent perturbation theory for lattice models with fermions}

%
\author{V.K. Sazonov$^{a,b}$}
\affil{$^a$Institute of Physics, Department of Theoretical Physics, University of Graz, \\ Universit\"atsplatz 5, Graz, A-8010, Austria\\
$^b$Department of Theoretical Physics, St. Petersburg State
University, Uljanovskaja 1, \\St. Petersburg, Petrodvorez, 198504,
Russia\\vasily.sazonov@uni-graz.at, vasily.sazonov@gmail.com}

\maketitle

\begin{abstract}
The standard perturbation theory in QFT and lattice models leads to asymptotic
expansions. However, an appropriate regularization of the path or lattice integrals allows one
to construct convergent series with an infinite radius of the convergence. In the earlier
studies this approach was applied to the purely bosonic systems. 
Here, using bosonization, we develop the convergent perturbation theory 
for a toy lattice model with interacting fermionic and bosonic fields.
\end{abstract}

\section{Introduction}
One of the main tools for studying quantum field theories far away from the small coupling limit is the lattice Monte Carlo.
However, lattice simulations are not always applicable. For instance, in case of the complex action they fail
because of the sign problem \cite{deForcrand} and it becomes necessary to develop alternative methods.
Here we do only a first step and do not consider the models with the sign problem, but develop 
a new approach for lattice computations based on the convergent perturbative expansion. 

Application of the standard perturbation theory to 
QFTs and lattice models is restricted to the region of small values of coupling constants (expansion parameters). 
This is caused by the asymptotic character of appearing series \cite{guillou1990large}. 
Nevertheless, it is possible to regularize initial integrals and to get the
convergent expansions. One of the possible regularizations is based
on the direct cut-off of the large fluctuations of the fields \cite{Meurice2002, Meurice2004, Meurice2005, Meurice20052, Meurice2006}.
This method works well for one dimensional integrals, but in
multi-dimensional case it leads to the complicated analytically unsolvable integrals.
Alternative regularization scheme was developed in \cite{Belokurov1, Belokurov2, Belokurov3, BelSolSha97, BelSolSha99, BelokurovEng}.
\footnote{For the constructions of the convergent expansions without regularization see \cite{Shaverdyan1983, UshveridzeSuper, Nalimov, IvanovProc}.}
Its main advantage is that all loop integrals which appear in computations are exactly the same as in the standard perturbation theory.
In this paper we generalize the latter method of constructing the convergent perturbation theory and apply
it to a lattice model with interacting fermionic and bosonic fields, a toy model of lattice QED.

The standard perturbative expansions are asymptotic because of the incorrect 
interchange of orders of the integration and summation. 
Conditions under which the integration and summation are interchangeable, are given by the following
version of the Fubini's theorem, see \cite{kolmogorov1999elements}.
\begin{theorem}
Let $(a, b)$ is a finite or infinite interval, $u_n(t)$ is a sequence of continuous 
complex functions defined on $(a, b)$ and
\begin{itemize}
  \item $\sum_{n = 0}^\infty u_n(t)$ converges uniformly on every bounded interval in $(a, b)$,
  \item at least one of the following quantities defined with Lebesgue integrals is finite 
\[
  \int_a^b \left(\sum_{n = 0}^\infty |u_n(t)| \right) dt\,,~~~\sum_{n = 0}^\infty\int_a^b |u_n(t)| dt\,.
\]
\end{itemize}
Then 
\[
  \int_a^b \left(\sum_{n = 0}^\infty u_n(t) \right) dt = \sum_{n = 0}^\infty\int_a^b u_n(t) dt\,.
\]
\label{theorem1}
\end{theorem}

In Ref. \cite{Belokurov1} the convergent perturbation theory satisfying the theorem \ref{theorem1} was constructed 
for the multi-dimensional integrals in ${\bf{R}}^N$ space. The generalization to the path integrals 
with the trace-class operators in the Gaussian measure was obtained
in \cite{Belokurov2, Belokurov3}. In both cases the interaction part of integrals was restricted to the 
continuous polynomials $P(x) \geq 0$ with the even degree $2 m$. In Section \ref{sec:Nd} we present the construction \cite{Belokurov1} 
extended to the integrals of the type
\begin{equation}
I(g) = \int_\RN e^{-\|x\|^2- g P(x)} dx
\end{equation}
with the interaction represented by the bounded from below continuous polynomial
$P(x) \geq -M,~M \geq 0$ with the even degree $2 m$.
In Section \ref{sec:QED} using the bosonization of the fermion determinant \cite{Luescher1994, Bunk1995, Slavnov1996, Slavnov1996a, Slavnov1999, Slavnov2000, Bakeev1997} 
we generalize convergent perturbation theory to the models containing fermions.
We demonstrate our method on the example of toy lattice model which may be considered as a simplest
approximation to the lattice QED in the Lorentz gauge.

\section{N-dimensional integrals}
\label{sec:Nd}
We consider the integral of the form
\begin{equation}
I(\bg) = \int_\RN e^{-x_i K_{ij} x_j- P(x)} dx\,,
\label{intN}
\end{equation}
where $K_{ij}$ is the matrix with strictly positive eigenvalues
and $P(x)$ is a bounded from below polynomial of even degree $2m$.
The main goal of this section is to build up its power expansion satisfying
the theorem \ref{theorem1}.

The polynomial $P(x)$ is bounded from below, but can be negative. We define its minimal value as
$P_{\text{min}} \equiv -M$ and then add and subtract
$M$ in the argument of the exponential function in (\ref{intN})
\begin{eqnarray}
I(\bg) = e^M \int_\RN e^{-x_i K_{ij} x_j- [P(x) + M]} dx\,.
\label{Ndex2}
\end{eqnarray}
Now the polynomial $\widetilde{P}(x) = P(x) + M \geq 0$ and to construct converging perturbation
theory we apply the program developed in \cite{Belokurov1}.
Let us introduce the Fourier transform of $\exp(-r^{2m})$ function
\begin{equation}
  \varphi_m(\rho) = \frac{1}{2\pi}\intpmi e^{-i \rho r} e^{-r^{2m}} dr\,.
\label{vphm}
\end{equation}
Employing the definition (\ref{vphm}) and positivity of $\widetilde{P}(x)$ we have 
\begin{equation}
  e^{-\widetilde P(x)} = \intpmi \varphi_m(\rho) e^{i \rho\, \widetilde{P}^{\frac{1}{2m}}(x)} d\rho\,.
\end{equation}
The equation (\ref{Ndex2}) transforms to
\begin{equation}
I(\bg) = e^M \int_\RN e^{-x_i K_{ij} x_j} \left[\intpmi \varphi_m(\rho) e^{i \rho\, \widetilde{P}^{\frac{1}{2m}}(x)}\, d\rho \right] dx\,.
\label{intN2}
\end{equation}
At large $|\rho|$ the function $\varphi_m(\rho)$ obeys
\begin{equation}
  |\varphi_m(\rho)| \leq C \exp\left(-|\rho|^{1 + \frac{1}{2m}}\right)\,,
\end{equation}
see \cite{BelSolSha97}.
Then one can estimate the integrand in (\ref{intN2}) as
\begin{equation}
  \left|e^{-x_i K_{ij} x_j} \varphi_m(\rho)\, e^{i \rho\, \widetilde{P}^{\frac{1}{2m}}(x)}\right| 
  \leq C \exp\left(-|\rho|^{1 + \frac{1}{2m}} - x_i K_{ij} x_j\right)\,.
\end{equation}
Consequently, the integral (\ref{intN2}) is absolutely convergent in the Lebesgue sense and according 
to the Fubini's theorem it is possible to interchange the order of integrations
\begin{equation}
I(\bg) = e^M \intpmi \varphi_m(\rho) \left(\int_\RN e^{-x_i K_{ij} x_j}e^{i \rho\, \widetilde{P}^{\frac{1}{2m}}(x)} dx \right) d\rho\,.
\label{intN3}
\end{equation}
The integral $I(\bg)$ may be represented as a limit of the proper integral over $\rho$
\begin{equation}
  I(\bg) = \lim_{R\rightarrow \infty} J(\bg, R)\,,
\end{equation}
where
\begin{equation}
J(\bg, R) = e^M \int_{-R}^{R} \varphi_m(\rho) \left(\int_\RN e^{-x_i K_{ij} x_j}e^{i \rho\, \widetilde{P}^{\frac{1}{2m}}(x)} dx \right) d\rho\,.
\label{JintN}
\end{equation}
Expanding the function $e^{i \rho\, \widetilde{P}^{\frac{1}{2m}}(x)}$ we get
\begin{equation}
  J(\bg, R) = e^M \int_{-R}^{R} \varphi_m(\rho)\, 
  \left[\int_\RN e^{-x_i K_{ij} x_j} 
  \left(
    \sum_{n = 0}^\infty \frac{i^n\rho^n \widetilde{P}^{\frac{n}{2m}(x)}}{n!} 
  \right)
  dx\right] d\rho\,.
\label{JintNserIntern}
\end{equation}
In this case the conditions of the theorem \ref{theorem1} are satisfied and
\begin{eqnarray}
J(\bg, R) = e^M \sum_{n = 0}^{\infty} \int_{-R}^{R} \varphi_m(\rho)\, (i\rho)^n\, d\rho \, 
  \int_\RN \frac{\widetilde{P}^{\frac{n}{2m}}(x)}{n!}\, e^{-x_i K_{ij} x_j} dx
\label{JintNser}
\end{eqnarray}
is the absolutely convergent series. The integrals over $x$ and $\rho$
are factorized and
\begin{eqnarray}
J(\bg, R) = e^M \sum_{n = 0}^{\infty} \frac{A_{n}(m, R)}{n!}\, 
  \int_\RN \widetilde{P}^{\frac{n}{2m}}(x)\, e^{-x_i K_{ij} x_j} dx\,,
\label{JintNser2}
\end{eqnarray}
where
\begin{equation}
 A_n(m, R) = i^n\, \int_{-R}^{R} \varphi_m(\rho)\, \rho^n d\rho = 
 \frac{1}{\pi} \intpmi \left(\frac{d^n}{dr^n}\, e^{-r^{2m}}\right) \, \frac{\sin R r}{r}\, dr\,.
\end{equation}
The properties of the coefficients $A_n(m, R)$ were carefully studied in \cite{Belokurov1}.

To proceed with the calculation of the integral over $x$ from (\ref{JintNser2}), we rewrite it as
\begin{equation}
  \int_\RN \widetilde{P}^l(x) \widetilde{P}^{\frac{\kappa}{2m}}(x)\, e^{-x_i K_{ij} x_j} dx\,,
\label{intrep}
\end{equation}
where $l \in \mathbb{N}$ and $\kappa = 1,..,(2m-1)$.
Let $\lambda$ be a half of the smallest eigenvalue of $K$, then we add and subtract
$\lambda$ times identity matrix in the argument of the exponential function in (\ref{intrep})
\begin{equation}
  \int_\RN \widetilde{P}^l(x) \left(\widetilde{P}(x) e^{-\frac{2m}{\kappa} \lambda x_i \delta_{ij} x_j} \right)^{\frac{\kappa}{2m}}\, e^{-x_i (K_{ij} - \lambda \delta_{ij}) x_j} dx\,.
\label{intrep2}
\end{equation}
The function $y = \widetilde{P}(x) e^{-\frac{2m}{\kappa} \lambda x_i \delta_{ij} x_j}$ is positive and bounded for any $x$. 
Then there is such $a> 0$, that $0 \leq y \leq a$. According to the Weierstrass theorem
the function $y^{\kappa / (2m)}$ can be approximated by a finite degree polynomial with an arbitrary precision.
For each $\delta > 0$ there exists a polynomial
\begin{equation}
\nonumber
a_0 + a_1 y +...+ a_s y^s\,,
\end{equation}
that for all $y \in [0, a]$
\[
  |y^{\kappa / (2m)} - a_0 - a_1 y -...- a_s y^s| < \delta\,.
\]
After the substitution of $y^{\kappa / (2m)}$ by its polynomial approximation the integral over $x$
reduces to a sum of the Gaussian integral moments
which can be easily taken.
This finishes the construction of convergent perturbation theory for the integral (\ref{intN}).

Finally, we remark that instead of the function $\varphi_m(\rho)$ one may use any function of the type
\begin{equation}
  \widetilde\varphi(\rho) = \frac{1}{2\pi}\intpmi e^{-i \rho r} e^{-r^k} dr\,,~~~\text{with even}~k \geq 2m\,.
\label{vpha}
\end{equation}
The corresponding regularized integral then looks as
\begin{equation}
  J(\bg, R) = e^M \int_{-R}^{R} \widetilde{\varphi}(\rho)\, 
  \left[\int_\RN e^{-x_i K_{ij} x_j} e^{i\rho \widetilde{P}^{\frac{1}{k}}(x)} dx\right] d\rho
\label{1dexJdefk}
\end{equation}
and still satisfies the theorem \ref{theorem1}, since 
\begin{equation}
  J(\bg, R) = e^M \int_{-R}^{R} \widetilde{\varphi}(\rho)\, 
  \left[\int_\RN e^{-x_i K_{ij} x_j} e^{|i\rho \widetilde{P}^{\frac{1}{k}}(x)|} dx\right] d\rho
\label{1dexJdefkabs}
\end{equation}
is finite, because $e^{-x_i K_{ij} x_j}$ decays at infinity
sufficiently faster then $e^{|i\rho \widetilde{P}^{\frac{1}{k}}(x)|}$ grows.

\section{Fermions}
\label{sec:QED}
The method presented in Section \ref{sec:Nd} was developed only for bosonic integrals.
Employing the bosonization of the fermion determinant \cite{Luescher1994, Bunk1995, Slavnov1996, Slavnov1996a, Slavnov1999, Slavnov2000, Bakeev1997}
we extend the convergent perturbation theory to the models with fermions. The procedure which is proposed remains
similar for any theory where the fermion contribution can be reduced to the strictly positive determinant. Here we focus on the toy model with interacting
bosons and fermions which we derive as a simplest lattice approximation to QED in the Lorentz gauge.
The Euclidean action of the continuum QED with two degenerate flavors in the Lorentz gauge is
\begin{equation}
S_{QED} = \int dx~ \big[\frac{1}{4} (F_{\mu\nu})^2 + \frac{1}{2\alpha}(\partial_\mu A_\mu)^2 -
\sum_{f = 1, 2} \bar\psi_f (i \gamma_\mu D_\mu + m) \psi_f \big]\,.
\label{LQED}
\end{equation}
Integration over fermions leads to the determinant
\begin{equation}
  \det (D + m)^2 = \det(\gamma_5(D + m)\gamma_5(D + m)) = \det(-D^2 + m^2) \equiv \det (B^2 + m^2)\,,
\label{det}
\end{equation}
where $B\equiv \gamma_5 D$ and $D = (\gamma_\mu\partial_\mu + i e \gamma_\mu A_\mu)$.
To obtain the toy model of lattice QED we apply a naive discretization 
to the gauge part of the action (\ref{LQED}) and to the operator $B$.
For this we define $x$ variable on the finite $4$-dimensional lattice with spacing $a$ and volume $V$
and substitute all derivatives by the finite difference approximations.
Then, the gauge action is given by
\begin{eqnarray}
\nonumber 
S_{gauge} = a^4 \sum_{x} {\cal L}_{gauge}\\
\nonumber
=a^4 \sum_{x}\{ \frac{1}{4} 
\left(\frac{A_\nu(x + a\hat\mu) - A_\nu(x)}{a} - \frac{A_\mu(x + a\hat\nu) - A_\mu(x)}{a}\right)^2\\
+
\frac{1}{2\alpha}\left(\sum_\mu \frac{A_\mu(x + a\hat\mu) - A_\mu(x)}{a}\right)^2\}\,.
\label{SGd}
\end{eqnarray}
The action of the operator $B$ on some spinor $\psi$ transforms to
\begin{equation}
  \hat{B}(x)\psi(x) = \gamma_5 \big[\gamma_\mu \frac{\psi(x + a\hat{\mu}) - \psi(x)}{a} + i e A_\mu(x) \psi(x)\big]\,.
\label{Bd}
\end{equation}
The equations (\ref{SGd}) and (\ref{Bd}) define our model of lattice QED.

Following \cite{Bakeev1997} we represent the determinant $\det(\hat{B}^2 + m^2)$ as a result of the
integration over the five dimensional bosonic fields $\phi_{n}(x)$
\begin{eqnarray}
\det (\hat{B}^2 + m^2) = \lim_{L\rightarrow \infty\,,\,b\rightarrow 0} \prod_{x} \int [d\phi^*_{n}(x)] [d\phi_{n}(x)] [d\xi^*(x)] [d\xi(x)]
\exp\{
-a^4 b \sum_{n = 0}^{N - 1} \sum_{x} {\cal L}_{matter}\}\,,
\label{detBos}
\end{eqnarray}
\begin{eqnarray}
\nonumber
{\cal L}_{matter} =
\Big(
\frac{\phi^*_{n+1}(x) - \phi^*_{n}(x)}{b} 
- 
i \hat{B}(x) \phi^*_{n}(x)
\Big)
\Big(
\frac{\phi_{n+1}(x) - \phi_{n}(x)}{b} 
+ 
i \hat{B}(x) \phi_{n}(x)
\Big)\\ 
+\sqrt{L}\big[\xi^*(x)(m + i\hat{B}(x))\phi_n(x) 
+ \text{h.c.}\big]
+\frac{1}{2 m} \xi^*(x)\xi(x)
\,.
\end{eqnarray}
Here $L$ - is the length of the additional dimension, $N$ is a number of sites and $b$ is a lattice spacing in this dimension,
\begin{equation}
  L = Nb\,,~~~0 \leq n < N\,,~~~b \|\hat{B}\| \ll 1\,.
\end{equation}
The fields $\phi_{n}(x)$ are five dimensional, but have the same spinorial structure as $(3+1)$-dimensional fermionic fields $\bar\psi(x), \psi(x)$.
We impose free boundary conditions in the additional dimension
\begin{equation}
  \phi_{n} = 0\,,~~~n < 0\,,~~~ n \geq N\,.
\end{equation}
The fields $\xi(x)$ are $(3+1)$-dimensional and bosonic.
Substituting the expression for $\hat{B}(x)$ we obtain
\begin{eqnarray}
\nonumber
{\cal L}_{matter} &=&
\left(
\frac{\phi^*_{n+1}(x) - \phi^*_{n}(x)}{b} 
- 
i \gamma_5\gamma_\mu \frac{\phi^*_{n}(x + a\hat{\mu}) - \phi^*_{n}(x)}{a} +  e\gamma_5 \gamma_\mu A_\mu(x) \phi^*_{n}(x)
\right)\\ 
\nonumber
&\cdot&
\left(
\frac{\phi_{n+1}(x) - \phi_{n}(x)}{b} 
+ 
i \gamma_5\gamma_\mu \frac{\phi_{n}(x + a\hat{\mu}) - \phi_{n}(x)}{a} -  e\gamma_5 \gamma_\mu A_\mu(x) \phi_{n}(x)
\right)\\ 
\nonumber
&+&\sqrt{L}\Big[m\xi^*(x)\phi_n(x) 
+ \xi^*(x)\big(i \gamma_5\gamma_\mu \frac{\phi_{n}(x + a\hat{\mu}) - \phi_{n}(x)}{a} -  e\gamma_5 \gamma_\mu A_\mu(x) \phi_{n}(x)\big) 
+ \text{h.c.}\Big]\\
\nonumber
&+&\frac{1}{2 m} \xi^*(x)\xi(x)
\,.\\
\label{LDF}
\end{eqnarray}
The Gaussian part of the Lagrangian (\ref{LDF}) is
\begin{eqnarray}
\nonumber
{\cal L}_{0} &=&
\left(
\frac{\phi^*_{n+1}(x) - \phi^*_{n}(x)}{b} 
- 
i \gamma_5\gamma_\mu \frac{\phi^*_{n}(x + a\hat{\mu}) - \phi^*_{n}(x)}{a}
\right)\\ 
\nonumber
&\cdot&
\left(
\frac{\phi_{n+1}(x) - \phi_{n}(x)}{b} 
+ 
i \gamma_5\gamma_\mu \frac{\phi_{n}(x + a\hat{\mu}) - \phi_{n}(x)}{a}
\right)\\ 
\nonumber
&+&\sqrt{L}\Big[m\xi^*(x)\phi_n(x) 
+ i \xi^*(x) \gamma_5\gamma_\mu \frac{\phi_{n}(x + a\hat{\mu}) - \phi_{n}(x)}{a}
+ \text{h.c.}\Big]\\
\nonumber
&+&\frac{1}{2 m} \xi^*(x)\xi(x)
\,.\\
\end{eqnarray}
The interaction part of ${\cal L}_{matter}$
\begin{eqnarray}
\nonumber
{\cal L}_{int} &=&
\left(
\frac{\phi^*_{n+1}(x) - \phi^*_{n}(x)}{b} 
- 
i \gamma_5\gamma_\mu \frac{\phi^*_{n}(x + a\hat{\mu}) - \phi^*_{n}(x)}{a}
\right)
\left(
 - e \gamma_5 \gamma_\mu A_\mu(x) \phi_{n}(x)
\right)\\
\nonumber
&+&
\left(
e \gamma_5 \gamma_\mu A_\mu(x) \phi^*_{n}(x)
\right)
\left(
\frac{\phi_{n+1}(x) - \phi_{n}(x)}{b} 
+ 
i \gamma_5\gamma_\mu \frac{\phi_{n}(x + a\hat{\mu}) - \phi_{n}(x)}{a}
\right)\\
\nonumber
&+& e^2 \gamma_\mu A_\mu(x) \phi^*_{n}(x) \gamma_\mu A_\mu(x) \phi_{n}(x)\\
\nonumber
&+&\sqrt{L}
\Big[ 
-e\xi^*(x) \gamma_5 \gamma_\mu A_\mu(x) \phi_{n}(x) 
+ \text{h.c.}\Big]\\
\label{LDFint}
\end{eqnarray}
is not bounded from below. This is caused by the separation of the Gaussian and interaction parts 
of the Lagrangian. To apply the method from the previous section we represent ${\cal L}_{int}$ as
a limit of the bounded function. There are infinitely many ways to do this, here we use the most trivial scheme 
\begin{equation}
\nonumber
  {\cal L}_{int}(x, n) = \lim_{\epsilon \rightarrow 0} \widetilde{\cal L}_{int}(x, n, \epsilon)\,,
\end{equation}
where
\begin{eqnarray}
\nonumber
  \widetilde{\cal L}_{int}(x, n, \epsilon) &=& {\cal L}_{int}(x, n) 
  + \epsilon\big[(A_\mu(x) A_\mu(x))^2 + (\phi^*_{n}(x) \phi_{n}(x))^2 \\ 
  &+& (\phi^*_{n + 1}(x) \phi_{n +1}(x))^2 + (\xi^*(x) \xi(x))^2\big]
\end{eqnarray}
is bounded from below.

The fields $\phi$, $\xi$ are complex, however, one can interpret them in terms of their real components
$\phi = c + i d$, $\xi = f + i h$.
Then the partition function of the model (\ref{SGd}), (\ref{Bd}) may be written
in the form of the equation (\ref{intN})
\begin{eqnarray}
\nonumber
{\cal Z} &=& \lim_{L\rightarrow \infty\,,\,b\rightarrow 0}~ \lim_{\epsilon\rightarrow 0} \prod_{x, n} \int [d A_x] [d\bar\phi_{x,n}] [d\phi_{x,n}] [d\bar\xi_x] [d\xi_x]\\
&\cdot&
\exp\{-a^4 b \sum_{n = 0}^{N-1} \sum_{x} [\frac{1}{L} {\cal L}_{gauge} + {\cal L}_{0} + \widetilde{\cal L}_{int}(x, n, \epsilon)]\}\,.
\label{QEDPS}
\end{eqnarray}
Note that the limit $\epsilon\rightarrow 0$ is not a priory interchangeable with other limits and should be taken first.

Let $M$ be an absolute value of the minimum of $\widetilde{\cal L}_{int}(x, n, \epsilon)$, we define $\Lambda_{int}(x, n, \epsilon) \equiv \widetilde{\cal L}_{int}(x, n, \epsilon) + M \geq 0$
and regularize (\ref{QEDPS}) as in the previous section
\begin{eqnarray}
\nonumber
{\cal Z}(R) &=& e^{\widetilde V M} \lim_{L\rightarrow \infty\,,\,b\rightarrow 0}~ \lim_{\epsilon\rightarrow 0} \prod_{x, n} \int_{-R}^R\, \varphi(\rho)
\Big[ \int [d A_x] [d\bar\phi_{x,n}] [d\phi_{x,n}] [d\bar\xi_x] [d\xi_x]\\
\nonumber
&\cdot&
\exp\{-a^4 b \sum_{n = 0}^{N-1} \sum_{x} [\frac{1}{L} {\cal L}_{gauge} 
+ {\cal L}_{0}]\}\\
&\cdot&
\exp\{i \rho \Big[ a^4 b \sum_{n = 0}^{N-1} \sum_{x}  \Lambda_{int}(x, n)\Big]^{\frac{1}{4}}\}\Big] d\rho\,,
\label{QEDPSR}
\end{eqnarray}
where $\widetilde V = V L$ is the volume of the five dimensional lattice.
The Taylor expansion of the second exponential function gives convergent perturbation theory for toy model of lattice QED. 

\section{Conclusions}
\label{sec:Concl}
The series of the standard perturbation theory for path and lattice integrals are asymptotic. This happens
because of the illegal interchange of the summation and integration. 
However, it is possible to regularize integrals in such a way, 
that all conditions required for the interchanging of the summation and integration will be satisfied.
This gives an approach to QFT and lattice computations valid at any arbitrary values of the coupling constants.

In this work we presented the method for constructing the convergent perturbation theory for integrals
with interactions bounded from below. Employing the bosonization of the fermion determinant we extended this
method to the model of lattice QED. Recently, a bosonization of the complex actions was proposed \cite{nonHBos}.
Together with the convergent perturbation theory this opens new way to avoid sign problem.

\section{Acknowledgments}
I acknowledge A. Alexandrova and C. Boutillier for discussions.
This work was supported by the Austrian AMS organization and by 
the Austrian Science Fund FWF Grant Nr. I 1452-N27.



\label{Bibliography} 
\bibliographystyle{unsrt}

\bibliography{bibliography}

\end{document}